\documentclass[12pt]{article}
\usepackage{graphicx,amssymb,epsfig,epsf} 
 \usepackage{ulem}
\usepackage[usenames]{color}
\usepackage{rotating}
\usepackage{epsfig}
\usepackage{slashed}
\usepackage{amsmath}
\usepackage{pstricks}
\usepackage{color}

\begin{document}

\begin{flushright}
   { OSU-HEP-13-09\
   }\\
\end{flushright}

\vskip 2pt

\begin{center}
{\large \bf Parallel Universe, Dark Matter and Invisible Higgs Decays }\\
\vskip 20pt
{
Shreyashi Chakdar$^{a}$\footnote{chakdar@okstate.edu}, {Kirtiman Ghosh$^{a}$\footnote{kirti.gh@gmail.com}}},
 S. Nandi$^{a}$\footnote{s.nandi@okstate.edu} 
   \\
\vskip 10pt
{Department of Physics and Oklahoma Center for High Energy Physics,\\
Oklahoma State University, Stillwater, OK 74078-3072, USA.}\\
\vskip 10pt

\end{center}

\vskip 5pt
\abstract
{The existence of the dark matter with amount about five times the ordinary matter is now well established experimentally. There are now many candidates for this dark matter. However, dark matter could be just like the ordinary matter in a parallel universe. If both universes are described by a non-abelian gauge symmetries, then there will be no kinetic mixing between the ordinary photon and the dark photon, and the dark proton, dark electron and the corresponding dark nuclei, belonging to the parallel universe, will be stable. If the strong coupling constant, $(\alpha_s)_{dark}$ in the parallel universe is five times that of $\alpha_s$, then the dark proton will be about five time heavier, explaining why the dark matter is five times the ordinary matter. However, the two sectors will still interact via the Higgs boson of the two sectors. This will lead to the existence of a second light Higss boson, just like the Standard Model Higgs boson. This gives rise to the invisible decay modes of the Higgs boson which can be tested at the LHC, and the proposed ILC.
}

\pagebreak

\section{Introduction}

Symmetry seems to play an important role in the classification and interactions of the elementary particles. The Standard Model (SM) based on the gauge symmetry $SU(3)_C \times SU(2)_L \times U_Y(1)$ has been extremely successful in describing all experimental results so far to a precision less than one percent.
The final ingredient of the SM, namely the Higgs boson, has finally been observed at the LHC \cite{ATLAS}. However, SM is unable to explain why the charges of the elementary particle are quantized because of the presence of $U(1)_Y$. This was remedied by enlarging the $SU(3)_C$ symmetry to $SU(4)_C$ with the lepton number as the fourth color,(or grand unifying all three interaction in SM in $SU(5)$ \cite{GG}
or $SO(10)$ \cite{gfm}).

 SM also has no candidate for the dark matter whose existence is now well established experimentally \cite{dm}. Many extensions of the SM models, such as models with weakly interacting massive particles (WIMP) can explain the dark matter \cite{dm}. The most poplar examples are the lightest stable particles in supersymmetry \cite{dm}, or the lightest Kaluza-Klein partcle in extra dimensions \cite{kkdm}. Of course, axion \cite{ww} is also a good candidate for dark matter. Several experiments are ongoing to detect signals of dark matter in the laboratory. However, it is possible that the dark matter is just the analogue of ordinary matter belonging to a parallel universe. Such a parallel universe  naturally appears in the superstring theory with the $E_8 \times E'_8$ gauge symmetry before compactification \cite{chsw}. Parallel universe in which the gauge symmetry is just the replication of our ordinary universe, i,e the gauge symmetry in the parallel universe being $SU(3)'\times SU(2)'\times U(1)'$ has also been considered \cite{fv}. If the particles analogous to the proton and neutron in the parallel universe is about five times heavier than the proton and neutron of our universe, then that will naturally explain why the dark matter of the universe is about five times the ordinary matter. This can be easily arranged by assuming strong coupling constant square$/4\pi, \alpha'_s$ is about five times larger than the QCD $\alpha_s$. Thus, in this work, we assume that the two universe where the electroweak sector is exactly symmetric, whereas the corresponding couplings in the strong sector are different, explaining why the dark matter is larger than the ordinary matter. Also, we assume that both universes are described by non-abelian gauge symmetry so that the kinetic mixing  between the photon ($\gamma$) and the parallel photon ($\gamma '$) is forbidden.
We also assume that post-inflationary reheating in the two worlds are different, and the the parallel universe is colder than our universe \cite {mohapatra}. This makes it possible to maintain the successful prediction of the big bang nucleosynthesis, though the number of degrees of freedom is increased from the usual SM of 10.75 at the time of nucleosynthesis  due the extra light degrees of freedom (due to the $\gamma ', e'$ and three $\nu '$s). 

In this work, we explore the LHC implications of this scenario due to the mixing among the Higgs bosons in the two electroweak sectors. Such a mixing, which is allowed by the gauge symmetry, will mix the lightest Higgs bosons of our universe ($h_1$) and the lightest Higgs boson of the parallel universe ($h_2$), which we will call the dark Higgs. One of the corresponding mass eigenstates, $h_{SM}$ we identify with the observed Higgs boson with mass of $125$ GeV. The other mass eigenstate, which we denote by $h_{DS}$, the dark Higgs, will also have a mass in the electroweak scale. Due to the mixing effects, both Higgs will decay to the kinematically allowed  modes in our universe and as well as to the modes of the dark universe. One particularly interesting scenario is when the two Higgs bosons are very close in mass, say within 4 GeV so that the LHC can not resolve it \cite {ATLASHgaga}
. However, this scenario will lead to the invisible decay modes\cite{ATLASINV} . The existence of such invisible decay modes can be established at the LHC when sufficient data accumulates. (The current upper limit on the invisible decay branching ratio of the observed Higgs at the LHC is $0.65$). At the proposed future International Linear Collider (ILC) \cite{ILC}, the existence of such invisible modes can be easily established, and the model can be tested in much more detail.

\section{Model and the Formalism}

The gauge symmetry we propose for our work is $ SU(4)_C \times SU(2)_L \times SU(2)_R $ for our universe, and $ SU(4)'_C \times SU(2)'_L \times SU(2)'_R $ for the parallel universe. Note that we choose this non-abelian symmetry not only to explain charge quantization (as in Pati-Salam model \cite{Pati:1974yy}), but also to avoid the kinetic mixing of $\gamma$ and $\gamma '$ as would be allowed in the Standard Model. All the elementary particles belong to the representations of this symmetry group and their interactions are governed by this symmetry. The 21 gauge bosons belong to the adjoint representations $(15,1,1)$, $(1,3,1)$, $(1,1,3)$. $(15,1,1)$ contain the 8 usual colored gluons, 6 lepto-quark gauge bosons $(X, \bar{X})$, and one $(B-L)$ gauge boson \cite{mm}. $(1,3,1)$ contain the 3 left handed weak gauge bosons, while $(1,1,3)$ contain the 3 right handed weak gauge bosons. The parallel universe contains the corresponding parallel gauge bosons. However, so far as the gauge interactions are concerned, we do not assume that the coupling for $SU(4)$ and $SU(4)'$ interactions are the same, but strong coupling in the parallel universe is larger in order to account for the $p'$ (proton of the parallel universe) mass to be about five times larger than the proton. For the electroweak sector, we assume the exact symmetry between our universe and the parallel universe.

 The fermions belong to the fundamental representations $( 4, 2,1) + (4,1,2)$. The  4 represent three color of quarks and the lepton number as the 4th color, $(2,1)$ and $(1,2)$ represent the left and right handed doublets. The forty eight Weyl fermions belonging to three generations may be represented by the matrix

\begin{equation}
{\begin{pmatrix}

{\begin{pmatrix} u \\ d \end{pmatrix}}_1 &
 {\begin{pmatrix} u \\ d \end{pmatrix}}_2 & {\begin{pmatrix} u \\ d \end{pmatrix}}_3
& {\begin{pmatrix} \nu_e \\ e \end{pmatrix}}_4\\

{\begin{pmatrix} c \\ s \end{pmatrix}}_1 &
 {\begin{pmatrix} c \\ s \end{pmatrix}}_2 & {\begin{pmatrix} c \\ s \end{pmatrix}}_3
& {\begin{pmatrix} \nu_{\mu} \\ \mu \end{pmatrix}}_4\\

{\begin{pmatrix} t \\ b \end{pmatrix}}_1 &
 {\begin{pmatrix} t \\ b \end{pmatrix}}_2 & {\begin{pmatrix} t \\ b \end{pmatrix}}_3
& {\begin{pmatrix} \nu_{\tau} \\ \tau \end{pmatrix}}_4\\
\end{pmatrix}}_{L, R}.
\end{equation} 

We have similar fermion representations for the parallel universe, denoted by primes.

 The model has 3 gauge  coupling constants: $g_4$ for $SU(4)$ color which we will identify with the strong coupling constant of our universe,  $g'_4$ for $SU(4)'$ color of the parallel universe, and $g$ for $SU(2)_L$ and $SU(2)_R$, and corresponding electroweak couplings for the parallel universe ($g_L = g_R = g'_L = g'_R = g$ (we assume that the gauge couplings of the electroweak sectors of the two universe are the same).

\subsection{Symmetry breaking}
 $SU(4)$ color symmetry is spontaneously broken to $SU(3)_C \times U(1)_{B-L}$ in the usual Pati-Salam way using  the Higgs fields $(15, 1,1)$ at a scale $V_c$.  The most stringent limit on the scale of this symmetry breaking comes from the upper limit of the rare decay mode $K_L \rightarrow \mu e$ \cite{KL}. 
$SU(2)_L \times SU(2)_R \times U(1)_{B-L}$ can be broken to the SM using the Higgs representations  $(1,2,1)$ and $1,1,2)$ at a scale $V_{LR}$. Alternatively, one can use the Higgs multiplets $(1.3,1)$ and $(1,1,3)$ if we want to generate the light neutrino masses at the observed scale.  Finally the remaining symmetry is broken to the $U(1)_{EM}$ using the Higgs bi-doublet $(1,2,2)$ as in the left-right model. The $(15, 2, 2)$ Higgs multiplet  could also be added to eliminate unwanted mass relations among the charged fermions. Similar Higgs representations are used to break the symmetry in the parallel universe to $U'(1)_{EM}$. A study of the Higgs potential shows that there exist a parameter space  where only one neutral Higgs in the bi-doublet remains light, and becomes very similar to the SM Higgs in our universe \cite{Senjanovic}. All other Higgs fields become very heavy compared to the EW scale. Similar is true in the parallel universe. The symmetry of the Higgs fields in   the EW sector between our universe and the parallel universe will make the two  electroweak VEV's the same. Thus the mixing terms between the two bi-doublets (one in our universe and one in the parallel universe) then leads to mixing between the two remaining SM like Higgs fields. The resulting mass terms for the remaining two light Higgs fields  can be written as 
$m_{VS}^2 h_{1}^2 + m_{DS}^2 h_{2}^2 + 2 \lambda v_{VS} v_{DS}h_1 h_2$, (where $v_{VS}$ and $v_{DS}$ are the electroweak symmetric breaking scale in the visible sector and dark sector respectively)  from which the two mass eigenstates and the mixing can be calculated. The implications for this is when the two light Higgses are very close in mass (within about 4 GeV, which LHC can not resolve) leads to the invisible decay of the observed Higgs boson. Below we discuss the  phenomenological implications for this scenario at the LHC, and briefly at the proposed ILC \cite{ILC}.


\section{Phenomenological Implications}

\begin{figure}
\begin{center}
\includegraphics[width=10.5 cm,height=8.5cm]{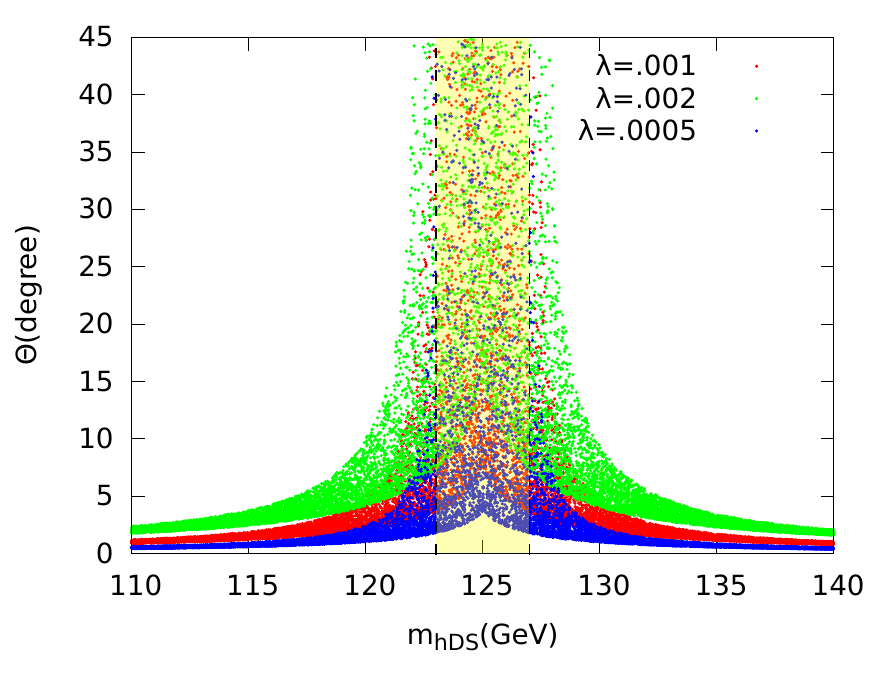}
\end{center}
\caption{Scatter plot of dark Higgs mass vs mixing angle for different values of $\lambda$. The SM-like Higgs mass is kept fixed in the range between 123 to 127 GeV denoted by the shaded region in the plot.}
\label{mixing}
\end{figure}

In the framework of this model, interaction between fermions and/or gauge bosons of dark sector and visible sector (the SM particles) are forbidden by the gauge symmetry. However, quartic Higgs interactions of the form $\lambda (H_{VS}^{\dag}H_{VS})(H_{DS}^{\dag}H_{DS})$ (where $H_{VS}$ and $H_{DS}$ symbols denote the Higgs fields in the visible sector and dark sector respectively) are allowed by the gauge symmetry and gives rise to mixing between the Higgses of dark and visible sector. The mixing between the lightest Higgses of dark sector and visible sector gives rise to interesting phenomenological implications at the collider experiments. In this section, we will discuss the phenomenological implications of the lightest dark and visible neutral Higgs mixing ($h_1$ and $h_2$). As discussed in the previous section, the bi-linear terms involving the lightest visible sector (denoted by $h_{1}$) and dark sector (denoted by $h_2$) Higgses in the scalar potential are given by,
\begin{equation}
{\cal L}_{Scalar} \supset m_{VS}^2h_1^2 + m_{DS}^2h_2^2 + 2 \lambda v_{VS} v_{DS} h_1 h_2
\end{equation}     
where, $v_{VS}$ and $v_{DS}$ are the electroweak symmetric breaking scale in the visible sector and dark sector respectively. In our analysis, we have assumed  the both $v_{VS}$ and $v_{DS}$ are equal to the SM electroweak symmetry breaking scale $v_{SM} \sim 250$ GeV. $m_{VS}$, $m_{DS}$ and $\lambda$ are the free parameters in the theory and the masses ($m_{h_1^{(p)}}$ and $m_{h_2^{(p)}}$) and mixing between physical light Higgs states (denoted by $h_1^{(p)}$ and $h_{2}^{(p)}$) are determined by these parameters:   
\begin{eqnarray}
h_{1}^{(p)}&=&{\rm cos}\theta ~h_1 + {\rm sin}\theta ~h_2, \nonumber\\
h_{2}^{(p)}&=&-{\rm sin}\theta ~h_1 + {\rm cos}\theta ~h_2,
\end{eqnarray}
where the masses and the mixing angle of these physical states are given by,
\begin{eqnarray}
m_{h_1^{(p)},h_2^{(p)}}^2&=&\frac{1}{2}[(m_{VS}^2+m_{DS}^2)\mp\sqrt{(m_{VS}^2-m_{DS}^2)^2+4\lambda^2v_{VS}^2v_{DS}^2}]\nonumber\\
{\rm tan}2\theta &=&\frac{2\lambda ~v_{VS}~v_{DS}}{m_{DS}^2-m_{VS}^2}.
\end{eqnarray}
 In the framework of this model, we have two light physical neutral Higgs ($h_1^{(p)}$ and $h_{2}^{(p)}$) states. Out of these two Higgs states, we define the SM like Higgs $h_{SM}$ is the state which is dominantly $h_1$-like, i.e., if ${\rm cos}\theta > {\rm sin}\theta$ then $h_{SM}=h_{1}^{(p)}$ and vice versa. The other Higgs is denoted as dark Higgs ($h_{DS}$).  Since ATLAS and CMS collaborations have already detected a SM like Higgs boson with mass about 125 GeV, we only studied the scenario where the mass of $h_{SM}$ is between 123 to 127 GeV. Before going into the details of collider implication of visible sector and dark sector Higgs mixing, it is important to understand the correlation between the mixing and mass of the dark Higgs ($m_{h_{DS}}$). To understand the correlation, for few fixed values of $\lambda$, we have scanned the $m_{VS}-m_{DS}$ parameter space.
 We have only considered the points which gives rise to a $h_{SM}$ in the mass range between 123 GeV to 127 GeV. For these points, the resulting dark Higgs masses ($m_{h_{DS}}$) and mixing ($\theta$) are plotted in Fig.~\ref{mixing}.
The scatter plot in Fig.~\ref{mixing} shows that large mixing in the visible and dark sector is possible only when the dark Higgs mass is near 125 GeV i.e., near the mass of SM like Higgs boson. It is important to note that the LHC is a proton-proton collider, i.e., LHC collides the visible sector particles only. Therefore, the production cross-section of dark Higgs at the LHC is proportional to the square of the visible sector Higgs component in $h_{DS}$. Therefore, in order to detect the signature of dark Higgs at the collider experiments, we must have significant mixing between the visible and dark sector Higgses. And Fig.~\ref{mixing} shows that significant mixing arises only when dark Higgs and SM like Higgs are nearly degenerate in mass. Therefore, in this article, we studied the phenomenology of two nearly degenerate Higgs bosons with mass about 125 GeV.

\subsection{Interactions and Decays of light Higgses}

In the present model, two light Higgs physical states ($h_1^{(p)}$ and $h_2^{(p)}$) result from the mixing of  visible sector and dark sector light Higgs weak eigenstate $h_1$ and $h_2$ respectively. Visible sector light Higgs weak eigenstates, $h_1$ interacts only with the visible sector fermions ($f$) via Yukawa interactions and gauge bosons ($V$) via gauge interactions. Whereas the dark sector light Higgs weak eigenstate interacts only with the dark fermions $f_D$ and dark gauge bosons $V_D$. However, as a result of mixing, the physical light Higgses interact with both the visible particles and dark particles and thus, they can be produced at the Large Hadron Collider(LHC) experiment. The coupling of the physical states $h_1^{(p)}$ and $h_2^{(p)}$ with the visible as well as dark fermions and gauge bosons can be written as a product of corresponding SM coupling and sine or cosine of the mixing angle. As a result the production cross sections of $h_1^{(p)}$ and $h_2^{(p)}$ and decay widths into visible as well as dark particles can be computed in terms of the SM Higgs production cross-sections/decay widths and the mixing angle. For example, total $h_1^{(p)}$ production cross section at the LHC is given by $\sigma_{SM}{\rm cos}^2\theta$, where $\sigma_{SM}$ is the production cross-section of the SM Higgs with equal mass. Similarly, the decay widths of $h_{1}^{(p)} (h_{2}^{(p)})$ into visible and dark sector fermions are given by $\Gamma_{SM}^{H\to f\bar f} {\rm cos}^2\theta$ ($\Gamma_{SM}^{H\to f\bar f} {\rm sin}^2\theta$) and $\Gamma_{SM}^{H\to f\bar f} {\rm sin}^2\theta$ ($\Gamma_{SM}^{H\to f\bar f} {\rm sin}^2\theta$) respectively, where $\Gamma_{SM}^{H\to f\bar f}$ is the decay width of the SM Higgs into fermions. It is important to note that since the QCD coupling in the dark sector is about $5$ times
larger than the QCD coupling in the visible sector, the Higgs coupling with dark gluon in this model is enhanced by a factor about $5$.\\
In this analysis we are considering both the higgs states in the mass range between $123-127$ GeV. Here we present the expressions for $\mu= \sigma/\sigma_{SM}$ and total  $\sigma\times BR_{invible}$ for present model,
\begin{eqnarray}
\mu &=&\frac{(\sigma_{h1}{cos}^4\theta BR_{h1}/(1 + 24 BR_{h1}^{gg}{sin}^2\theta)) + (\sigma_{h2}{sin}^4\theta BR_{h2}/(1 + 24 BR_{h2}^{gg} {cos}^2\theta)) }{\sigma_{SM}*BR}\nonumber\\
\sigma \times BR_{inv} &=&\frac{\sigma_{h1}{cos}^2\theta {sin}^2\theta (BR_{h1}^{inv} + 25BR_{h1}^{gg})}{1+ 24BR_{h1}^{gg} {sin}^2\theta} + \frac{\sigma_{h2}{cos}^2\theta {sin}^2\theta (BR_{h2}^{inv} + 25BR_{h2}^{gg})}{1+ 24BR_{h2}^{gg} {cos}^2\theta}\
\end{eqnarray}
where $\sigma_{h1}$ corresponds to Standard Model Higgs production cross-section at mass of $h_{1}^{(p)}$ and $\sigma_{h2}$ corresponds to Standard Model production cross-section at mass of $h_{2}^{(p)}$ (see Table \ref{production cross section}) and $BR_{h1}$ and $BR_{h2}$ corresponds to Branching ratios of Higgs boson at mass $h_{1}^{(p)}$ and $h_{2}^{(p)}$ respectively(see Table \ref{Decay Branching Ratio1}). For calculating the $\mu$ values in present model we have used Branching Ratios of $H \rightarrow WW \rightarrow l \nu l \nu$ and $H \rightarrow \gamma \gamma$ channels(see Table \ref{Decay Branching Ratio}).
\subsection{Data used in Collider Analysis}

In this section, we discuss the collider phenomenology of invisible Higgs Decays. Before going into
the details of the collider prediction, we first need to study the constraints on the parameter space
coming from present Standard Model predictions and experimental data. The Higgs mass eigenstates of $h_{SM}$ and $h_{DS}$ will be produced in Colliders through the top loop as top quark has Standard Model couplings to the $h_{SM}$ mass eigen state.  
The Higgs, which comprises of both $h_1$ and $h_2$ eigen states, will then decay in both the Standard Model decay modes along with Dark sector decay modes. We will perceive these dark sector decay modes as enhancement in the invisible Branching Fraction of the Higgs.\

 We first discuss the different constraints on the mixing angle $\theta$ between the two eigenstates coming from experimental data of $H \rightarrow WW \rightarrow l \nu l \nu$ and $H \rightarrow \gamma \gamma$ channels. Along with these experimental data in Higgs decays in different modes, we have also taken into account constraints on the mixing angle parameter space coming from the ATLAS search for the invisible decays of a $125$ GeV Higgs Boson produced in association with a Z boson \cite{ATLASINV}. \ 
\begin{table}
\begin{center}
\begin{tabular}{|c|c|c|c|c|}
\hline
Mass of Higgs(GeV) & $\sigma_{ggf}$ & $\sigma_{ttH}$ &$\sigma_{VBF}$ &$\sigma_{Vh}$ \\
\hline
\hline
123 & 20.15 & 1.608 & 1.15 & 0.1366 \\
\hline
124 & 19.83 & 1.595 & 1.12 & 0.1334 \\
\hline
125 & 19.52 & 1.578 & 1.09 & 0.1302 \\
\hline
126 & 19.22 & 1.568 & 1.06 & 0.1271 \\
\hline
127 & 18.92 & 1.552  & 1.03 & 0.1241  \\
\hline
\end {tabular}
\end{center}
\vspace{0.2cm}
\caption{ Standard Model production cross section (pb) in different channels for $E_{CM}$ = 8 TeV.} 
\label{production cross section}
\end{table}

\begin{table}
\begin{center}
\begin{tabular}{|c|c|c|c|c|c|}
\hline
Mass of Higgs(GeV) & BR($H {\rightarrow}WW$) & BR($H {\rightarrow}ZZ$) & BR($H {\rightarrow}{\gamma\gamma}$) & BR($H {\rightarrow} gg $)& BR($H {\rightarrow}ff$) \\
\hline
\hline
123 & 0.183 & $2.18\times 10^{-2}$ & $2.27 \times 10^{-3}$  & $8.71\times 10^{-2}$  & 0.687 \\
\hline
124 & 0.199 & $2.41\times 10^{-2}$ & $2.27\times 10^{-3}$ & $8.65\times 10^{-2}$ &  0.687 \\
\hline
125 & 0.215 & $2.64\times 10^{-2}$ & $2.28\times 10^{-3}$ & $8.57\times 10^{-2}$ & 0.670 \\
\hline
126 & 0.231 & $2.89\times 10^{-2}$ & $2.28\times 10^{-3}$ & $8.48\times 10^{-2}$ & 0.651 \\
\hline
127 & 0.248 & $3.15\times 10^{-2}$  & $2.27\times 10^{-3}$ & $8.37\times 10^{-2}$  & 0.633 \\
\hline
\end {tabular}
\end{center}
\vspace{0.2cm}
\caption{ Standard Model Decay Branching Ratio in different channels.} 
\label{Decay Branching Ratio1}
\end{table}

\begin{table}
\begin{center}
\begin{tabular}{|c|c|c|}
\hline
Channels for Higgs Decay & $\mu$ value by ATLAS & $\mu$ value by CMS \\
\hline
\hline
$H \rightarrow WW \rightarrow l \nu l \nu$  & $1.01 \pm 0.31$ & $0.76 \pm 0.21$ \\
\hline
 $H \rightarrow  \gamma \gamma$ & $1.65 \pm 0.24(stat) ^ {+0.25}_{-0.18}(syst) $  & $0.78 \pm 0.27$ \\

\hline
\end {tabular}
\end{center}
\vspace{0.2cm}
\caption{ Experimental values of best fit signal strength $\mu = \sigma/\sigma_{SM}$ at $E_{CM}$ = 8 TeV.} 
\label{Decay Branching Ratio}
\end{table}

The Standard Model production cross-sections in different channels (such as gluon-gluon fusion, ttH, vector boson fusion and vector boson (both W boson and Z boson) in association with a Higgs boson) at $E_{CM}$ = 8 TeV and Decay Branching ratios in different channels (such as $H {\rightarrow}WW$, $H {\rightarrow}ZZ$,$H {\rightarrow}{\gamma\gamma}$,$H {\rightarrow} gg$,$H {\rightarrow}ff$)  has been given by ATLAS collaboration in reference \cite{SM production} \cite{SM decay}. We have used these cross-sections and branching ratios in different channels in our analysis.
The relevant cross-sections and branching ratios used for our analysis are presented in Table \ref{production cross section} and Table \ref{Decay Branching Ratio1} respectively. We have taken the mass range between $123 - 127$ GeV which is the interesting parameter space for our analysis.\

In Table \ref{Decay Branching Ratio} we present the results of the different experimental searches in the $H \rightarrow WW \rightarrow l \nu l \nu$ channel by ATLAS collaborations \cite{ATLASWW} and CMS collaboration \cite{CMSWW} and in $H \rightarrow  \gamma \gamma$ channel by ATLAS collaborations\cite{ATLASGG} CMS collaborations\cite{CMSGG} .

\subsection{Bounds on Mixing Angle}
\begin{figure}
\begin{center}
\vspace*{-2.0cm}
\includegraphics[width=10.5cm,height=8.5cm]{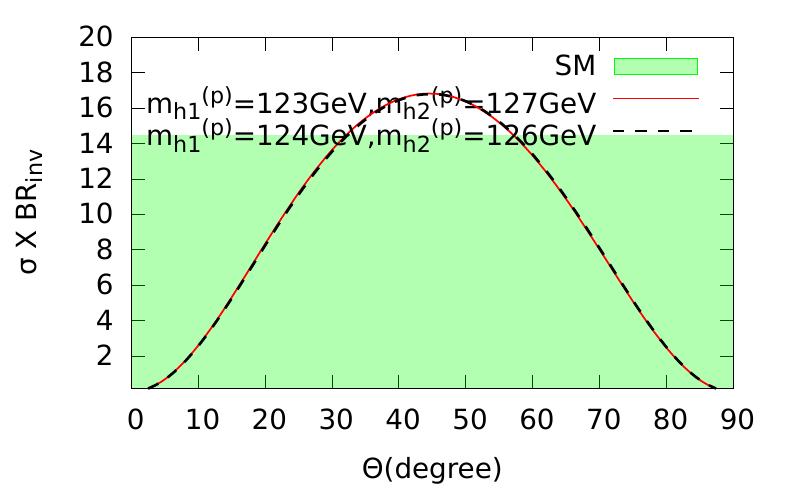}
\caption{Decay rate in invisible channels in present model as a function of mixing angle $\theta$. The shaded regions correspond to SM allowed values for $\sigma \times BR_{inv}$.} 
\label{fig:decayrate}
\end{center}
\end{figure}

In this section we use the data that we presented in the previous section to constrain the mixing angle parameter space.
In Fig \ref{fig:decayrate}, we present the total invisible decay rate i.e $\sigma \times BR$ in the invisible channel vs the mixing angle $\theta$ for
  $ m_{h1}^{(p)} = 123$ GeV  and $ m_{h2}^{(p)} = 127$ GeV ($m_{h1}^{(p)} = 124 GeV$ and $m_{h2}^{(p)} = 126 GeV)$ . ATLAS collaboration has searched for the invisible decay of higgs boson in Z H production channel at $E_{CM} = 8 TeV$. In absence of any significant deviation of data from the Standard Model background prediction, ATLAS collaboration has set an upper limit of $65\%$ on the invisible decay branching of a SM higgs boson of mass $125$ GeV \cite{ATLASINV}. Assuming $\sigma_{total}$ = $22.32$ pb Higgs cross-section at $125$ GeV (see Table \ref{production cross section}),  $65\%$ upper limit on invisible decay branching ratio corresponds to $14.5$ pb upper limit on the invisible Higgs decay rate. This limit is shown in the shaded green region in Fig \ref{fig:decayrate}. It can be seen from the plot that present model is consistent with ATLAS experimental data for $\theta < 33^{o} $ and $\theta > 58^{o}$ in the parameter space region. 

\begin{figure}
\begin{center}
\vspace*{-2.0cm}
\includegraphics[width=10.5cm,height=8.5cm]{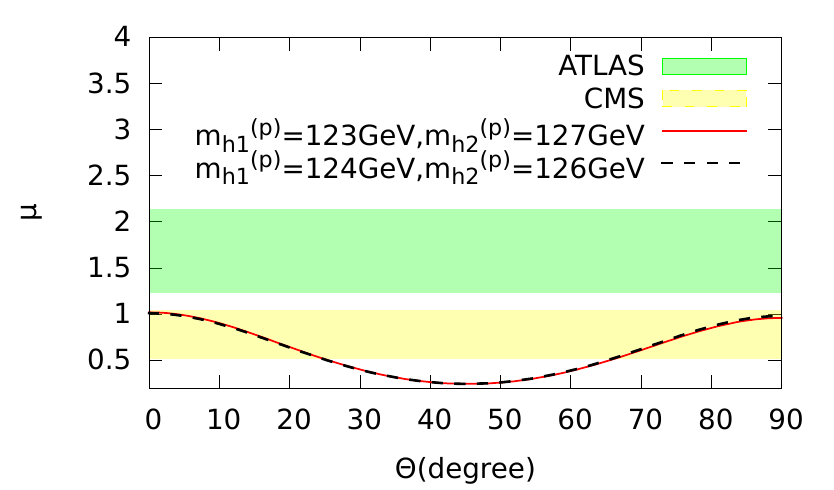}
\caption{Higgs decaying into diphoton rate in present model as a function of mixing angle $\theta$. The shaded regions again correspond to ATLAS and CMS allowed $\mu = \sigma/\sigma_{SM}$ values.} 
\label{fig:H_gaga}
\end{center}
\end{figure} 

\begin{figure}
\begin{center}
\vspace*{-2.0cm}
\includegraphics[width=10.5cm,height=8.5cm]{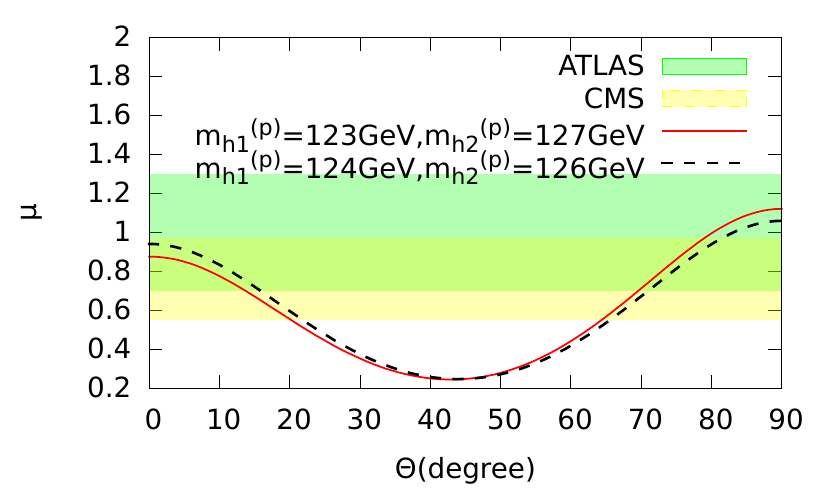}
\caption{$H \rightarrow WW \rightarrow l \nu l \nu$ rate in present model as a function of mixing angle $\theta$. The shaded regions correspond to ATLAS and CMS allowed $\mu = \sigma/\sigma_{SM}$ values.} 
\label{fig:H_WW}
\end{center}
\end{figure}

In Fig. \ref{fig:H_gaga} we have presented a plot of $\mu = \sigma/\sigma_{SM}$ in the $H \rightarrow \gamma \gamma$ channel as a function of the  mixing angle $\theta$. The plot shows prediction in present model for $m_{h1}^{(p)} = 123$ GeV  and $ m_{h2}^{(p)} = 127$ GeV $(m_{h1}^{(p)} = 124$ GeV and $m_{h2}^{(p)} = 126$ GeV) mass values.The yellow shaded region corresponds for allowed region by CMS collaboration and green shaded region is allowed region for ATLAS collaboration in this channel. It can be seen from the plot that CMS allowed region is consistent for all $\theta$'s for the present model,but present model is not consistent with ATLAS allowed region for any values of $\theta$. We point out that  $H \rightarrow  \gamma \gamma$ data for ATLAS, is well above the SM expectation. If the present model is realized by  Nature, with the accumulation of more data with higher luminosities at the Large Hadron Collider(LHC) the $H \rightarrow  \gamma \gamma$ branching ratio measured by ATLAS experiment should should come down significantly from present experimental value of $1.65 \pm 0.24(stat) ^ {+0.25}_{-0.18}(syst) $. Our model is consistent with the lower $\mu$ value of $0.78 \pm 0.27$  for $H \rightarrow  \gamma \gamma$ as measured by the CMS experiment for the whole parameter of the parameter space.

In Fig.~\ref{fig:H_WW} we present a plot of $\mu = \sigma/\sigma_{SM}$ in the $H \rightarrow WW \rightarrow l \nu l \nu$ channel with mixing angle $\theta$. Two curves for $ m_{h1}^{(p)} = 123$ GeV  and $ m_{h2}^{(p)} = 127$ GeV $(m_{h1}^{(p)} = 124$ GeV and $m_{h2}^{(p)} = 126$ GeV) present the prediction for present model. The yellow shaded region corresponds for allowed region by CMS collaboration and green shaded region is for allowed region by ATLAS collaboration in this channel. It can be seen from the plot that ATLAS allowed region is consistent with present model for $\theta < 13 (16)^{o} $ and $\theta > 70 (71)^{o} $ region in the parameter space. It can also be seen that present model is also consistent with CMS allowed region for $\theta < 20 (23)^{o}$ and $\theta > 65 (66)^{o}$ parameter space. It is  interesting to note that the prediction curves for the present model with mass values of $ m_{h1}^{(p)} = 123$ GeV  and $ m_{h2}^{(p)} = 127$ GeV $(m_{h1}^{(p)} = 124$ GeV and $m_{h2}^{(p)} = 126$ GeV) are not symmetric. It can be understood by taking into the fact that in low $\theta$ region $ m_{h1}^{(p)}$ is SM like. As $ m_{h1}^{(p)}$ is lower than $ m_{h2}^{(p)}$ for both curves, the cross-section $\times$ Branching ratio is smaller in lower $\theta$ region. Whereas for high $\theta$ region $ m_{h2}^{(p)}$ is SM like and as it is heavier than $ m_{h1}^{(p)}$ for both curves the cross section $\times$ Branching Ratio is higher in this region,which makes the curves non-symmetric.\\
This present analysis in the $H \rightarrow WW \rightarrow l \nu l \nu$ channel gives the most stringent constraint of $\theta < 13 (16)^{o} $ and $\theta > 70 (71)^{o} $ on the parameter space for the mixing angle $\theta$ taking into account all the constraints coming from analysis in $\sigma \times BR_{invisible}$, $H \rightarrow \gamma \gamma$ and $H \rightarrow WW \rightarrow l \nu l \nu$ channels. From this analysis in different channels it is certain that there is still plenty of parameter space available for the present model taking into account all the known experimental constraints at the LHC.\\
We would also like to comment that in a linear collider like the proposed International Linear Collider(ILC) this analysis can be done without any ambiguity about the resolution of the two Higgs in the close range of $4GeV$. In a $e+e-$ collider the Higgs will be produced in association with a Z boson and from the mass recoil of the Z boson the peak resolution of the Higgs boson can be measured in the limit of $40$ MeV \cite{ILC}. So from linear colliders we will be able to tell for sure if there are two Higgs bosons in the comparable mass range between ($123-127$GeV), which is not possible in this precision from Hadron Collider like LHC.  

\section{Summary and Conclusions}

Motivated by the fact that the dark matter is about five  times the ordinary matter, we have proposed that the dark matter can just be like the ordinary matter in a parallel universe with the corresponding strong coupling constant, $\alpha^{'}_s$ about five times the strong coupling, $\alpha_s$ of our universe. The parallel universe needs to be much colder than our universe to keep the successful prediction for the big bang nucleosynthesis. We have used the non-abelian Pati-Salam gauge symmetry for both universe to have the charge quantization, as well as, to avoid any kinetic mixing between the photon of our universe and the parallel universe. However, the two universes will be connected via the electroweak Higgs bosons of the two universes. If the electroweak sector of the two universes are symmetric, the lightest Higgs bosons of the two universes will mix. In particular, if these two Higgses mix significantly, and their masses are close (say within 4 GeV), LHC will not be able to resolve if it is observing one Higgs or two Higgses. However, each Higgs will decay to the particles of our universe as well as to the corresponding particles of the the parallel universe. This leads to the invisible decays  of the observed Higgs boson (or bosons). We have used all the available experimental data at the LHC to set constraint on this mixing angle, and find that in can be as large as $16^{o}$. If the mixing angle is not very small, LHC will be able to infer the existence of such invisible decays when sufficient data accumulates. (The current limit on the invisible branching ratio from the LHC data is
$<65\%$). We also find that the cross section times the branching ratio for Higgs to $\gamma \gamma$ channel is fully consistent with our model as measured by the CMS collaboration, but not by the ATLAS collaboration. The results by the ATLAS collaboration for this channel has to come down if our model is realized by nature. Our  proposal of two Higgses around $125$ GeV , and significant invisible decay fraction  can  easily be tested in the proposed ILC where peak resolution of the Higgs boson can be measured to about $40$ MeV.

{\bf Acknowledgment:} We thank K. S. Babu for some important comments and him and S. Gottlieb for useful discussions. The work of SC, KG and SN was supported in part by the U.S. Department of Energy
Grant Number DE-SC0010108.


\end{document}